\documentclass[12pt,onecolumn]{IEEEtran}

\usepackage[utf8]{inputenc} 
\usepackage[T1]{fontenc}
\usepackage{url}              
\usepackage{cite}             

\usepackage[cmex10]{amsmath}  
\usepackage{mleftright}       
\mleftright                   

\usepackage{graphicx}         
\usepackage{booktabs}         

\usepackage{float}
\usepackage{caption}
\usepackage[utf8]{inputenc} 
\usepackage[T1]{fontenc}
\usepackage{url}
\usepackage{ifthen}
\usepackage{graphicx}
\usepackage{amsthm}
\usepackage{amsfonts}
\usepackage{cite}
\usepackage{multicol}
\usepackage{color}
\usepackage{amssymb}
\usepackage[cmex10]{amsmath} 
\usepackage{stfloats}
\usepackage{enumitem}
\newlist{steps}{enumerate}{1}
\setlist[steps, 1]{label = Step \arabic*:}

\newcommand{\un}[1]{\underline{#1}}
\newcommand{\bb}[1]{\mathbb{#1}}

\newtheorem{defn}{Definition}
\newtheorem{thm}{{\cal T}heorem}
\newtheorem{cor}{Corollary}
\newtheorem{prop}{Proposition}
\newtheorem{lem}{Lemma}
\newtheorem{conj}{Conjecture}
\newtheorem{constr}{Construction}
\newtheorem{note}{Note}

\newtheorem{remark}{Remark}
\newcommand{\bit}{\begin{itemize}}
	\newcommand{\eit}{\end{itemize}}
\newcommand{\bcor}{\begin{cor}}
	\newcommand{\ecor}{\end{cor}}
\newcommand{\beq}{\begin{equation}}
	\newcommand{\eeq}{\end{equation}}
\newcommand{\beqn}{\begin{equation}}
	\newcommand{\eeqn}{\end{equation}}
\newcommand{\bea}{\begin{eqnarray}}
	\newcommand{\eea}{\end{eqnarray}}
\newcommand{\bean}{\begin{eqnarray*}}
	\newcommand{\eean}{\end{eqnarray*}}
\newcommand{\ben}{\begin{enumerate}}
	\newcommand{\een}{\end{enumerate}}
\newcommand{\bdefn}{\begin{defn}}
	\newcommand{\edefn}{\end{defn}}
\newcommand{\bnote}{\begin{note}}
	\newcommand{\enote}{\end{note}}
\newcommand{\bprop}{\begin{prop}}
	\newcommand{\eprop}{\end{prop}}
\newcommand{\blem}{\begin{lem}}
	\newcommand{\elem}{\end{lem}}
\newcommand{\bthm}{\begin{thm}}
	\newcommand{\ethm}{\end{thm}}
\newcommand{\bconj}{\begin{conj}}
	\newcommand{\econj}{\end{conj}}
\newcommand{\bconstr}{\begin{constr}}
	\newcommand{\econstr}{\end{constr}}
\newcommand{\bpf}{\begin{proof}}
	\newcommand{\epf}{\end{proof}}
\newcommand{\brem}{\begin{remark}}
	\newcommand{\rem}{\end{remark}}

\newcommand{\R}{R_{\textnormal{opt}}}

\newcommand{\cc}{\mathcal{C}}

\newcommand{\baln}{\begin{align*}}
	\newcommand{\ealn}{\end{align*}}
\newcommand{\bal}{\begin{align}}
	\newcommand{\eal}{\end{align}}

\usepackage{titlesec}

\IEEEoverridecommandlockouts
\begin{document}
	\title{On Streaming Codes for Simultaneously Correcting Burst and Random Erasures}
\author{Shobhit Bhatnagar$^\star$, Biswadip      Chakraborty$^\dagger$ and P. Vijay Kumar$^\star$\\
	$^\star$Department of Electrical Communication Engineering, IISc Bangalore \\
	$^\dagger$Qualcomm, India\\
	\{shobhitb97, chakrabortibiswadip, pvk1729\}@gmail.com
		\thanks{This research is supported by the "Next Generation Wireless Research and Standardization on 5G and Beyond" project funded by MeitY and SERB Grant No. CRG/2021/008479. The work of Shobhit Bhatnagar is supported by a Qualcomm Innovation Fellowship India 2021.}
	}
	\maketitle
\begin{abstract}
	Streaming codes are packet-level codes that recover dropped packets within a strict decoding-delay constraint. We study streaming codes over a sliding-window (SW) channel model which admits only those erasure patterns which allow either a single burst erasure of $\le b$ packets along with $\le e$ random packet erasures, or else, $\le a$ random packet erasures, in any sliding-window of $w$ time slots. We determine the optimal rate of a streaming code constructed via the popular diagonal embedding (DE) technique 
	over such a SW channel under delay constraint $\tau=(w-1)$ and provide an $O(w)$ field size code construction.
	For the case $e>1$, we show that it is not possible to significantly reduce this field size requirement, assuming the well-known MDS conjecture. We then provide a block code construction whose DE yields a streaming code achieving the rate derived above, over a field of size sub-linear in $w,$ for a family of parameters having $e=1.$ We show the field size optimality of this construction for some parameters, and near-optimality for others under a sparsity constraint. Additionally, we derive an upper-bound on the $d_{\text{min}}$ of a cyclic code and characterize cyclic codes which achieve this bound via their ability to simultaneously recover from burst and random erasures. 
\end{abstract}

\section{Introduction}
\label{sec:intro}
\begin{figure}
	\begin{center}
	{\scalebox{0.6}{\includegraphics{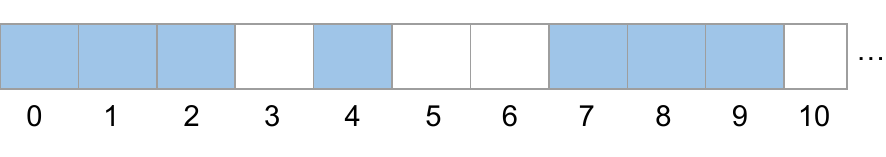}}} 
	\caption{An erasure pattern that is admissible in the $(2,(3,1),5)$-SW channel. Colored squares denote erasures.}
	\label{fig:erasure_pattern_fig}
	\end{center}
\end{figure}

Streaming codes are packet-level codes that recover erased or dropped packets within a strict decoding-delay constraint. Streaming codes were first studied in \cite{MartSunTIT04,MartTrotISIT07}, where the authors considered a decoding-delay constraint $\tau,$ and a sliding-window (SW) channel model that allowed a burst of $\le b$ packet erasures in any SW of length $w$. The authors derived an upper-bound on the rate of a streaming code for such a setting, and came up with a matching code construction that diagonally embeds codewords of a systematic scalar block code in the packet stream.
The model in \cite{MartSunTIT04,MartTrotISIT07} was then extended in \cite{BadrPatilKhistiTIT17} to one in which any SW of length $w$ can contain either a burst erasure of length $\le b$, or else $\le a$ random erasures. Such channel models serve as tractable approximations to the Gilbert-Elliot (GE) channel model. The optimal rate of a streaming code over such a SW channel model is known for all parameters, and can be achieved by diagonally embedding a carefully constructed systematic block code \cite{NikDeepPVK,KhistiExplicitCode}.
However, in \cite{Khisti_partialRecovery_burst_and_random} the authors showed that one can find a better approximation to a GE channel using SW channels that admit, in any SW of length $w,$ either a burst erasure of length $\le b$ {\emph{simultaneously}} with $\le e$ random erasures (where $e\ge1$), or else, a set of $\le a$ random erasures. We will call such a channel model an $(a,(b,e),w)$-SW channel (an example is illustrated in Fig. \ref{fig:erasure_pattern_fig}). We will refer to streaming codes over the $(a,(b,e),w)$-SW channel model under delay constraint $\tau$ as $(a,(b,e),w,\tau)$ streaming codes. It can be assumed \cite{BadrPatilKhistiTIT17,ShoBisPVK_burst_and_random} that the parameters of an $(a,(b,e),w,\tau)$ streaming code satisfy $\tau\ge(w-1)\geq (b+e)>a$. The authors in \cite{Khisti_partialRecovery_burst_and_random} remark that the case $e=1$ is the most frequent while approximating a Gilbert-Elliot channel, and provide a code construction for partial recovery of messages when $e=1$ and $w$ is large. In \cite{ShoBisPVK_burst_and_random} the authors derive an upper-bound on the rate of an $(a,(b,e),w,\tau)$ streaming code for the case $e<b,$ and provided a matching code construction for the case $e=(b-1)$. We refer the reader to \cite{RudRas_learning_augmented,RamBhaPVK_near_optimal_isit,Khisti_partialRecovery_burst_and_random,ShoBisPVK_burst_and_random,CloMed_multipath,Khisti_three_node_relay,RudRas_online_vs_offline,AdlCas,HagKriKhi_unequal, RLSC2,KriFacDom_3node, RamVajPVK_locally_recoverable_SC,RudRas_learning,Fra_delay_opt,Fra_wt,BadLuiKhi_burst_multicast,MahBadKhis_burst_rank_loss,VajRamNikPVK,VajRamNikPVK_simple_streaming_codes} for the study of streaming codes for various settings and channel models.

\emph{Our Contributions:} We determine the optimal rate of an $(a,(b,e),w,w-1)$-streaming code constructed via diagonal embedding of a systematic linear block code. We then show that this rate can be achieved via diagonal embedding of an MDS code requiring an $O(w)$ field size. Further, assuming the well-known MDS conjecture, we show that it is not possible to significantly reduce this field size requirement when $e>1.$ For the case $e=1,$ we provide a code construction for some families of parameters that requires a field size that is sub-linear in $w,$ as a special case of a more general construction of a block code that recovers from two burst erasures, where the length of one burst divides the length of the other. 
We show field size optimality of our construction for some parameters and near-optimality for others under a sparsity constraint.
Lastly, we upper-bound the $d_{\text{min}}$ of a cyclic code and characterize cyclic codes which achieve this bound via their ability to simultaneously recover from burst and random erasures. 

The organization of the paper is as follows. In Section \ref{sec:background_and_preliminaries} we introduce the notation and provide background on streaming codes. In Section \ref{sec:opt_rate} we determine the optimal rate of an $(a,(b,e),w,w-1)$ streaming code constructed by diagonal embedding of a systematic block code. In Section \ref{sec:const}, we provide our sub-linear field size code construction and show its field size optimality/near-optimality. Section \ref{sec:cyc_code_results} discusses cyclic codes.

\section{Background and Preliminaries}
\label{sec:background_and_preliminaries}
\subsection{Notation} 
Given integers $m,n$, we will use $[m:n]$ to denote the set $\{m,m+1,\dots,n\},$ and $m|n$ to denote that $m$ divides $n$. We will use the notation $l_{(n,m)}:=\lceil \frac{n}{m}\rceil.$
We will use $\un{0}_n$ to denote the zero vector of length $n,$ $\un{1}_n$ to denote the all-$1$ vector of length $n,$ and $\un{\un{0}}^{m\times n}$ to denote the all-zero matrix of size ${(m\times n)}$. We will use $I_n$ to denote the $(n\times n)$ identity matrix.
For a matrix $A\in\bb{F}^{m\times n}_q$, and $\mathcal{I}\subseteq[0:m-1],~\mathcal{J}\subseteq[0:n-1]$, ${A}(\mathcal{I},\mathcal{J})$ denotes the sub-matrix of $A$ consisting of the rows indexed by $\mathcal{I}$ and the columns indexed by $\mathcal{J}$. For $i\in[0:m-1]$ and $j\in[0:n-1]$, $A(i,:)$ and $A(:,j)$ denote the $i$-th row and the $j$-th column of ${A},$ respectively. Similarly, $A(\mathcal{I},:)$ and $A(:,\mathcal{J})$ denote the sub-matrices of $A$ consisting of the rows indexed by $\mathcal{I}$ and the columns indexed by $\mathcal{J}$, respectively.  We will at times use $A_j$ to denote the $j$-th column of $A$. For a vector $\un{v}=[v_0,v_1,\dots,v_{n-1}]^T\in\bb{F}_q^n,$ we will use $\text{supp}(\un{v})$ to denote the support of $\un{v},$ i.e., $\text{supp}(\un{v})=\{i~|~v_i\ne 0,~i\in[0:n-1]\}$. We will use $w_H(\un{v})$ to denote the Hamming weight of $\un{v},$ i.e., $w_H(\un{v})=|\text{supp}(\un{v})|$.

We will say that an $[n,k]$ code $\cc$ is systematic if its generator matrix $G$ is of the form $G=[I_k\ |\ P]\in\bb{F}^{k\times n}_q$. We will call a parity-check matrix $H$ of a systematic code to be in systematic form if it is of the form $H=[P'\ |\ I_{n-k}]\in\bb{F}^{(n-k)\times n}_q$.
For an $[n,k]$ code $\cc$, an erasure pattern
$\un{e}=[e_0,e_1,\dots,e_{n-1}]^T\in\{0,1\}^n$ signifies that $c_i,~i\in \text{supp}(\un{e})$ is erased, and $c_i,~i\in [0:n-1]\setminus \text{supp}(\un{e})$ is not erased, from a codeword $\un{c}^T=[c_0,c_1,\dots,c_{n-1}]\in\cc.$ 
We will call a code $\cc$ a $(b_1, b_2)$-code, where we will follow the convention $1\le b_2\le b_1,$ if it can recover from all erasure patterns that can be described as consisting of two erasure bursts, one of length $\le b_1,$ and the other of length $\le b_2.$

For a polynomial $p(X)=p_0+p_1X+\dots+p_kX^k$ of degree $k>0$, such that $p_0\ne0$, $Z(p(X))$ denotes the length of the longest run of consecutive zero coefficients in $p(X).$ Formally, $Z(p(X))=
\max\{(i_2-i_1)-1~|~p_i=0,~i\in[i_1+1:i_2-1],~0\le i_1< i_2\le k\}$ if at least one $p_i=0,~i\in[1:k-1],$ and $Z(p(X))=0$ otherwise.


\subsection{Streaming Codes}
\label{sec:streaming_code_framework}
\paragraph{Framework} We borrow the framework from \cite{MartSunTIT04}. At each time instant $t\in\bb{N}\bigcup\{0\},$ the encoder receives a message packet $\un{u}(t)=[u_1(t),u_2(t),\dots,u_{k}(t)]^T\in\bb{F}^k_q,$ and produces a coded packet $\un{x}(t)=[x_{1}(t),x_2(t),\dots,x_{n}(t)]^T\in\bb{F}^n_q.$ We will assume that $\un{u}(t)=\un{0}_k$ for $t<0.$ The encoder is causal, i.e., $\un{x}(t)$ is a function of prior message packets $\un{u}(0),\un{u}(1),\dots,\un{u}(t)$ only. The coded packet $\un{x}(t)$ is then transmitted over a packet erasure channel, so that at time $t,$ the receiver receives 
$$\un{y}(t)=\begin{cases}
	\un{x}(t),~\text{if}~\un{x}(t) \text{~is not erased,}\\
	\wedge,~\text{if}~\un{x}(t) \text{~is erased.}
\end{cases}$$
Here, we use the symbol $\wedge$ to denote an erased packet.
A delay-constrained decoder at the receiver must recover $\un{u}(t)$ using $\{\un{y}(0),\un{y}(1),\dots,\un{y}(t+\tau)\}$ only, where $\tau$ denotes the decoding-delay constraint, for all $t\ge0$. Packet-level codes satisfying this requirement are called streaming codes.
The rate of the streaming code is defined to be $\frac{k}{n}$.

\paragraph{Diagonal Embedding} In diagonal embedding (DE), codewords of an $[n,k]$ systematic block code $\cc$ are placed diagonally in the coded packet stream. See Fig. \ref{fi:DE} for an example.  More formally, if $G = [I_k \mid P]$ is a generator matrix of $\cc$, then a stream of $n$ packets in a rate $\frac{k}{n}$ streaming code constructed via DE of $\cc$ is of the form $[x_1(t),x_2(t+1),\dots,x_{n}(t+n-1)]=[u_1(t),u_2(t+1),\dots,u_{k}(t+k-1)]G$. Thus, DE translates packet erasures to codeword symbol erasures.
\begin{figure}
	\begin{center}
		\resizebox{.65\textwidth}{!} 
		{
			\begin{tabular}{ | c | c | c | c | c | c | c | c | }
				\hline
				{\color{red}\shortstack{$x_1(t)$}} & {\color{blue} \shortstack{$x_1(t+1)$}}&		{\color{black} \shortstack{$x_1(t+2)$}} & &  &  & & 
				\\ \hline
				& {\color{red} \shortstack{$x_2(t+1)$}}  & {\color{blue} \shortstack{$x_2(t+2)$}} & {\color{black} \shortstack{$x_2(t+3)$}}&  &  & &
				\\ \hline
				& & 	{\color{red}  \shortstack{$x_3(t+2)$}} & {\color{blue} \shortstack{$x_3(t+3)$}} & {\color{black} \shortstack{$x_3(t+4)$}}  & & &
				\\ \hline
				& & &	{\color{red} \shortstack{$x_4(t+3)$}}& {\color{blue} \shortstack{$x_4(t+4)$}} & {\color{black} \shortstack{$x_4(t+5)$}} & &
				\\ \hline
				& & & &	{\color{red} \shortstack{$x_5(t+4)$}}& {\color{blue} \shortstack{$x_5(t+5)$}} & {\color{black} \shortstack{$x_5(t+6)$}} &
				\\ \hline
				& & & & &	{\color{red} \shortstack{$x_6(t+5)$}}& {\color{blue} \shortstack{$x_6(t+6)$}} & {\color{black} \shortstack{$x_6(t+7)$}}  \\ \hline
			\end{tabular}
		}	
		\caption{DE of a $[6,3]$ systematic code $\mathcal{C}$. Each column is a coded packet and every diagonal of the same color is a codeword of $\mathcal{C}$.}
		\label{fi:DE}
	\end{center}
\end{figure}
%
\subsection{Preliminaries}
\label{subsec:preliminaries}
We will use the following straightforward lemma.
\blem
\label{lem:basic}
Let $\cc$ be an $[n,k]$ code and let $H$ be an $(n-k)\times n$ parity-check matrix of $\cc$. Let the coordinates in $E\subseteq[0:n-1]$ be erased from $\cc$. Then, $\cc$ can recover all the erased code symbols iff $\{H_i\mid i\in E\}$ is a linearly independent set.
\elem

An $[n,k,d]$ code $\cc$ is said to be Maximum Distance Separable (MDS) if it meets the well-known Singleton bound $d\le (n-k+1)$ with equality \cite[Chapter 1]{ecc_Mac_Slo}. The well-known MDS conjecture \cite[Chapter 11]{ecc_Mac_Slo} states that if $\cc$ is an $[n,k,d]$ MDS code over $\bb{F}_q,$ where $2\le k\le q,$ then $n\le(q+1),$ unless $q=2^m$ and $k=3$ or $k=(q-1),$ in which case $n\le(q+2).$ If $k>q,$ then $n\le(k+1).$

An $[n,k]$ code $\cc$ is said to be cyclic if $\forall \un{c}^T=[c_0,c_1,\dots,c_{n-1}]\in\cc,$ the vector $[c_{n-1},c_0,\dots,c_{n-2}]\in\cc$ \cite[Chapter 7]{ecc_Mac_Slo}. Let $\cc$ be an $[n,k]$ cyclic code over $\bb{F}_q$ and let $h(X)=h_0+ h_1 X+\dots+h_{k-1} X^{k-1}+h_k X^k$ be its reciprocal polynomial. Then $h_0\neq 0,~h_k\neq 0,$ and an ${(n-k)\times n}$ parity-check matrix of $\cc$ is given as 
\begin{align}
	\label{eq:pcm}
	H=
	\begin{bmatrix}
		h_0&h_1&\cdots&h_{k-1}&h_k&0&\cdots&0\\
		0&h_0&\cdots&h_{k-2}&h_{k-1}&h_k&\cdots&0\\
		\vdots&\vdots&\vdots&\vdots&\vdots&\vdots&\vdots&\vdots\\
		0&\cdots&0&h_0&\cdots&\cdots&h_{k-1}&h_k
	\end{bmatrix} . 
\end{align}

\section{Optimal Rate of $(a,(b,e),w,\tau=w-1)$ Streaming Codes Constructed via DE}
\label{sec:opt_rate}
The following theorem characterizes the optimal rate of an $(a,(b,e),w,\tau=w-1)$ streaming code constructed via DE.

\bthm
\label{thm:streaming_code_rate}
The optimal rate of an $(a,(b,e),w,\tau=w-1)$ streaming code constructed via DE is given by
\bea
\label{eq:streaming_code_bound}
\R =\frac{w-(b+e)}{w}.
\eea
\ethm
\bpf
The argument presented here is similar to one presented in \cite{LiKhistiGirod}. Suppose that $\cc$ is a systematic $[n,k]$ code, such that DE of $\cc$ results in an $(a,(b,e),w,\tau=w-1)$  streaming code. Since the code $\cc$ recovers from $(b+e)$ erasures, it follows that $(n-k)\ge (b+e)$. 

We will first consider the case $n<w$. Let $n=(w-\delta_1)$ and $k=(n-(b+e)-\delta_2)$, where $\delta_1>0$ and $\delta_2 \ge 0.$ It is easy to verify that in this case $\frac{k}{n}<\frac{w-(b+e)}{w}$. 
Now consider the case $n\ge w$, and consider the erasure pattern that erases the packet at time slot $0$ along with that last $(b+e-1)$ packets in the window $[0:w-1]$, i.e., the packets at time slots $t\in J=\{0\}\bigcup[w-(b+e)+1:w-1]$. Note that this erasure pattern is admissible in the $(a,(b,e),w)$-SW channel. Now consider the diagonally-embedded codeword $\un{c}=[c_0,c_1,\dots,c_{n-1}]=[x_1(0),x_2(1),\dots,x_{n}(n-1)]\in\cc.$ Since $\cc$ is a systematic code, the first $k$ code symbols $c_0,c_1,\dots,c_{k-1}$ correspond to message symbols. In order to recover the message symbol $m_0=c_0$ which has been erased, we must have that at least one of the code symbols in $\{c_i\mid i\in [0:w-1]\setminus J\}$ is a parity symbol. Thus, we must have $k\le w-(b+e)$. Combining this with the inequality $(n-k)\ge (b+e),$ we get $$\frac{k}{n}\le\frac{w-(b+e)}{w}.$$

Since any SW of length $w$ in the $(a,(b,e),w)$-SW channel can contain at most $(b+e)$ erasures, and the delay constraint is $(w-1),$ DE of a systematic $[w,w-(b+e)]$ MDS code yields an $(a,(b,e),w,\tau=w-1)$ streaming code of rate $\R$.
\epf

\begin{figure}
	\begin{center}
		{\scalebox{0.6}{\includegraphics{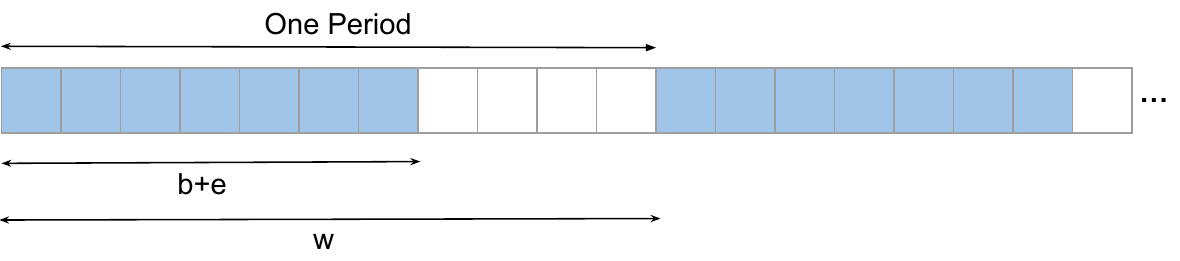}}} 
		\caption{A periodic erasure pattern admissible in the $(a,(b,e),w)$-SW channel when $e\ge (b-1)$. Colored squares denote erasures.}
		\label{fig:periodic_erasure_pattern_fig}
\end{center} 
\end{figure}

We remark here that in general, \eqref{eq:streaming_code_bound} is tighter than the upper-bound $\frac{w-a}{w-a+b+e+\frac{e}{m}}$ derived in \cite{ShoBisPVK_burst_and_random}, where $m=\lceil\frac{w-(b+e)}{b+e-a}\rceil$. Further, $\R$ is the optimal rate of any $(a,(b,e\ge(b-1)),w,\tau=w-1)$ streaming code. To see this, consider the erasure pattern in Fig. \ref{fig:periodic_erasure_pattern_fig}. By a direct extension of the argument in \cite{ShoBisPVK_burst_and_random}, this erasure pattern is admissible in the $(a,(b,e),w)$-SW channel. Each period of the erasure pattern is of length $w$ and has $w-(b+e)$ non-erased packets, yielding a rate upper-bound $\frac{w-(b+e)}{w}$, which is achievable from Theorem \ref{thm:streaming_code_rate}.


The arguments in the proof of Theorem \ref{thm:streaming_code_rate} imply that if DE of a systematic $[n,k]$ code yields an $(a,(b,e),w,\tau=w-1)$ streaming code achieving rate $\R,$ then $n=w,$ and $k=w-(b+e).$
The code constructed in Theorem \ref{thm:streaming_code_rate} requires a field of size $O(w)$. As we show in the next theorem, if $e>1,$ one can not do significantly better.

\bthm
Let $\cc$ be an $[n,n-(b+e)]$ code over $\bb{F}_q$ such that it can recover from any burst erasure of length $b$ along with any $e>1$ random erasures, where $n>(b+e+1)$. Then, $q\ge (n-b-2)$.
\ethm
\bpf
Let $H\in\bb{F}_q^{(b+e)\times n}$ be a parity-check matrix of $\cc$. Since $\cc$ recovers from the burst erasure that erases the first $b$ code symbols, $H([0:b+e-1],[0:b-1])$ must have full rank from Lemma \ref{lem:basic}. Hence, we can perform row operations on $H$ to get $H([0:b-1],[0:b-1])=I_b$ and $H([b:b+e-1],[0:b-1])=\un{\un{0}}^{e\times b}$. Now suppose that there exist $e$ distinct indices $i_0,i_1,\dots,i_{e-1}$ in $[b:n-1]$ such that the set of sub-columns $S=\{H_{i_j}([b:b+e-1])\mid j\in[0:e-1]\}$ is a linearly dependent set. Then we can combine $H_{i_j},~ j\in[0:e-1],$ to get a vector $\un{v}$ such that the last $e$ coordinates of $\un{v}$ are $0$. Thus, $\un{v}$ can be written as a linear combination of $H_0,H_1,\dots,H_{b-1},$ which implies that $\{H_j\mid j\in[0:b-1]\}\bigcup\{H_{i_j}\mid j\in[0:e-1]\}$ is a linearly dependent set. This can not be true since from the hypothesis of the theorem, $\cc$ recovers from any burst erasure of length $b$ along with any $e$ random erasures. Hence, any $e$ columns of $H([b:b+e-1],[b:n-1])$ must be linearly independent. In other words, $H([b:b+e-1],[b:n-1])$ must be a generator matrix of an $[n-b,e]$ MDS code. The claim follows from the MDS conjecture.
\epf



\section{A Sub-Linear Field Size Construction for A class of Parameters where $e=1$}
\label{sec:const}
We will first present a construction of a parity-check matrix $H$ of an $[n,n-(b_1+b_2)]$ $(b_1, b_2)$-code $\cc,$ where $b_2|b_1.$ This code is constructed over a field $\bb{F}_{q'},$ where $q'$ is the smallest power of a prime satisfying $q'\ge l_{(n,b_1)}.$ The steps of this construction, which we will call Construction I, are listed below:
\bit
\item \textit{Step 1:} Initialize $\tilde{H}\in\bb{F}_{q'}^{(b_1+b_2)\times (b_1l_{(n,b_1)})}$ to be the all-zero matrix.
\item \textit{Step 2:} Set  $\tilde{H}([0:b_1-1],[jb_1:(j+1)b_1-1])=I_{b_1},~j\in[0:l_{(n,b_1)}-1].$
\item \textit{Step 3:} Set $\tilde{H}([b_1:b_1+b_2-1],[0:b_1-1])=\un{\un{0}}^{b_2\times b_1}.$ 
\item \textit{Step 4:} Set $\tilde{H}([b_1:b_1+b_2-1],[jb_1:(j+1)b_1-1])=\alpha^{j-1}[I_{b_2}\ |\ I_{b_2}\ | \ \dots \ |\ I_{b_2}]\in\bb{F}_q^{b_2\times b_1},
~j\in[1:l_{(n,b_1)}-1],$ where $\alpha$ is any element of $\bb{F}_q$ whose order is $>(l_{(n,b_1)}-2).$
\item \textit{Step 5:} Set $H=\tilde{H}(:,[0:n-1]).$
\eit
For example, for $b_1=3,\ b_2=1$ and $n=8,$ Construction I yields the following parity-check matrix over $\bb{F}_3$: 
\bea
\label{eq:sublin_example}
H = \begin{bmatrix}
	1&0&0&1&0&0&1&0\\
	0&1&0&0&1&0&0&1\\
	0&0&1&0&0&1&0&0\\
	0&0&0&1&1&1&2&2\\
\end{bmatrix}.
\eea

In the theorem below, we prove that the code constructed via Construction I is a $(b_1, b_2)$-code.

\bthm
\label{thm:sub-linear_field_size}
The $[n,n-(b_1+b_2)]$ code $\cc$ constructed via Construction I is a $(b_1, b_2)$-code. Further, if $n>2b_1,$ this code has minimum distance $3,$ and if $n\le2b_1,$ it has minimum distance $4$.
\ethm

\bpf
We prove the theorem for an example case here. The proof for the general case is present in the appendix.
Consider the case $b_1=3,\ b_2=1$ and $n=8.$ The parity-check matrix of the code $\cc$ constructed via Construction I is given by \eqref{eq:sublin_example}.
We will now form a partition of the columns of $H$. For $s\in[0:2],$
let $H_i\in \mathcal{A}_s$ if $i=s\pmod{3},~i\in[0:7].$ Thus, $\mathcal{A}_0=\{H_0,H_3,H_6\}$, $\mathcal{A}_1=\{H_1,H_4,H_7\}$ and $\mathcal{A}_2=\{H_2,H_5\}$. Note that if $H_i\in \mathcal{A}_s,~\text{supp}(H_i)=\{s\}$ if $i\in[0:2],$ and $\text{supp}(H_i)=\{s,3\}$, if $i\in[3:7].$ It follows that for any $s_1,s_2\in[0:2]$ such that $s_1\neq s_2,$ any column in $\mathcal{A}_{s_1}$ is linearly independent of any column in $\mathcal{A}_{s_2
}.$ Further, by construction, any two columns $H_i$ and $H_j$ belonging to the same $\mathcal{A}_s$ are linearly independent. It follows that the support of any non-trivial linear combination of $H_i$ and $H_j$ is a non-empty subset of $\{s,3\}$. 

Now consider a burst erasure of length $3$ along with $1$ random erasure. Let $I$ be the set of indices corresponding to the burst erasure and let $j_0$ be the index of the random erasure. It is easy to check that $\{H_i\mid i\in I\}$ contains exactly one column from each $A_s$, $s\in [0:2]$, and thus $\{H_i\mid i\in I\}$ is a linearly independent set. Now, let $H_{j_0}\in \mathcal{A}_{s_0}.$ Then there exists exactly one element of $I$, say $i_0,$ such that $H_{j_0},H_{i_0}\in A_{s_0}$. One can check that the support of any non-zero vector in the linear span of $\{H_i\mid i\in I\setminus\{i_0\}\}$ contains at least one element of $[0:2]\setminus\{s_0\}.$ 
As mentioned before, $H_{j_0},H_{i_0}$ are linearly independent and the support of any non-trivial linear combination of $H_{j_0}$ and $H_{i_0}$ is a non-empty subset of $\{s_0,3\}$. It follows that $\{H_i\mid i\in I\bigcup\{j_0\}\}$ is a linearly independent set.

The minimum distance of $\cc$ is $3$ since $\mathcal{A}_0$ contains $3$ columns of $H$ which are linearly dependent.
\epf
We note here that the last $(b_1+b_2)$ columns of the matrix $H$ constructed via Construction I are linearly independent, and we can bring $H$ into systematic form via row operations. Thus, by setting $n=w,~b_1=b$ and $b_2=1,$ Construction I allows one to construct $(a,(b,1),w,\tau=w-1)$ streaming codes via DE, achieving rate $R_{\text{opt}},$ for any $w$ when $a=2,$ and $w\le2b$ when $a=3.$

\brem
If $b_1|n,$ following similar arguments as in the proof of Theorem \ref{thm:sub-linear_field_size}, one can show that the code constructed via Construction I can additionally recover from all erasure patterns which allow either the burst of length $b_1$ or the burst of length $b_2$ to be a wrap-around burst, i.e., a cyclic burst.
\rem

When $b_1=b_2=b,$ Construction I requires a field of size $l_{(n,b)}$. However, for the same parameters, a construction in \cite{LiKhistiGirod} (where the authors study delay-constrained decoding of multiple erasure bursts of the same length) requires a field of size $(l_{(n,b)}-1).$ As we will show below, if $b_1\neq b_2,$ then one can not get a field size improvement over that of Construction I in general.
\bthm
\label{thm:binary_to_ternary_optimality}
Construction I has the smallest field size of any $[n,n-(b+1)]$ $(b, 1)$-code, where $(b+1)<n\le3b.$ 
\ethm
\bpf
If $(b+1)<n\le 2b,$ Construction I yields a binary code. We will now consider the case $2b<n$ and show that one can not construct an $[n,n-(b+1)]$ $(b, 1)$-code over a field of size $<3,$ for any $b>1.$ We will prove this result by contradiction.
Let $H$ be the parity-check matrix of an $[n,n-(b+1)]$ $(b, 1)$-code over $\bb{F}_2,$ where $n>2b.$ Since $\cc$ is a $(b, 1)$-code, the first $(b+1)$ columns of $H$ must be linearly independent, and we can perform row operations so that $H(:,[0:b])=I_{b+1}.$ 
Now suppose that $\text{rank}(H([0:b]\setminus\{1\},[b+1:2b]))<b.$ Then, one can combine $H_i,\ i\in[b+1:2b],$ to get a vector $\un{v}=[v_0,v_1,\dots,v_b]^T$ such that $v_i=0,\ i\in[0:b]\setminus\{1\}.$ Since $H_1=I_{b+1}(:,1),$ it follows that $\{H_i\ |\ i\in[b+1:2b]\bigcup\{1\}\}$ is a linearly dependent set, and $\cc$ can not recover from the $(b, 1)$-burst whose support is given by $[b+1:2b]\bigcup\{1\}.$ Thus, we have shown that $(H([0:b]\setminus\{1\},[b+1:2b]))$ must be full-rank.
Now observe that for $j\in[b+1:n-1],$ we must have $H(0,j)=1$ since if $H(0,j)=0,$ then $H_j\in\textnormal{span}\langle\{H_i\ |\ i\in [1:b]\}\rangle.$ Similarly, for $j\in[b+1:n-1],$ we must have $H(b,j)=1,$ otherwise $H_j\in\textnormal{span}\langle\{H_i\ |\ i\in [0:b-1]\}\rangle.$ Thus, we get $H(0,[b+1:2b])=H(b,[b+1:2b])=\un{1}_b^T,$ which implies that $(H([0:b]\setminus\{1\},[b+1:2b]))$ is not full rank.
\epf

Analogous to Construction I, one can construct the parity-check matrix ${H}_{\text{bin}}$ of a binary $[n,n-(b_1+b_2\lceil\log_2(l_{(n,b_1)})\rceil)]$ $(b_1, b_2)$-code by replacing each distinct element in the last $b_2$ rows of the matrix $H$ constructed via Construction I by a distinct binary $\lceil\log_2(l_{(n,b_1)})\rceil$-tuple.
For example, for $b_1=3,~b_2=1$ and $n=8,$ we get
\bean
{H}_{\text{bin}} = \begin{bmatrix}
	1&0&0&1&0&0&1&0\\
	0&1&0&0&1&0&0&1\\
	0&0&1&0&0&1&0&0\\
	0&0&0&1&1&1&0&0\\
	0&0&0&0&0&0&1&1\\
\end{bmatrix},
\eean
where, in the last row of \eqref{eq:sublin_example}, we have replaced $0\in\mathbb{F}_3$ by $[0,0]^T,$ $1\in\mathbb{F}_3$ by $[1,0]^T$ and $2\in\mathbb{F}_3$ by $[0,1]^T.$
The proof that this construction works is analogous to the proof of Theorem \ref{thm:sub-linear_field_size}. Also, analogous to Theorem \ref{thm:binary_to_ternary_optimality}
one can show that if $(b+1)<n<3b$, as is the case in the above example, this construction is rate-optimal among the class of binary $(b, 1)$-codes of length $n$.

Let $\mathcal{H}(n,b)$ denote the set of parity-check matrices in systematic form of $[n,n-(b+1)]$ $(b, 1)$-codes over any finite field.
In the theorem below we show how one can use Construction I to obtain a sparsest member of $\mathcal{H}(n,b)$ (in terms of total number of entries which are zero) which has near-optimal field size under this sparsity constraint. We remark that sparsity is desirable in practice.

\bthm
Let $H$ be the parity-check matrix of the $[n,n-(b+1)]$ $(b, 1)$-code obtained via Construction I. Then, one can obtain a matrix $H'\in\mathcal{H}(n,b)$ by performing row operations on $H$ such that no other element of $\mathcal{H}(n,b)$ has lesser number of non-zero entries than $H'.$ Further, let $q^*$ be the smallest prime power such that there exists $\tilde{H}\in\bb{F}_{q^*}^{(b+1)\times n}$ such that $\tilde{H}\in\mathcal{H}(n,b)$ and $\tilde{H}$ has the same number of non-zero entries as $H'.$ Then
$q^*\geq l_{(n,b)}-1.$
\ethm
\bpf
Let $H'$ be the $(b+1)\times n$ parity-check matrix in systematic form of an $[n,n-(b+1)]$ $(b,1)$-code over $\bb{F}_q.$ Let $H''\in\bb{F}_q^{(b+1)\times n}$ be the matrix obtained by reversing the rows and columns of $H'$, i.e., $H''(i,j)=H(b-i,n-1-j),~i\in[0:b],~j\in[0:n-1].$ Then, $H''$ is of the form $H''=[I_{b+1}\mid P''].$ Clearly, one can obtain $H'$ by reversing the rows and columns of $H''.$ It is easy to see that $H''$ is also a parity-check matrix of an $[n,n-(b+1)]$ $(b,1)$-code $\cc$ over $\bb{F}_q$ (having the same number of non-zero entries as $H'$). For ease of notation, we will work with parity-check matrices of this form. We will first show that any column of $P''$ must have Hamming weight $\ge2.$ Then we will show that the first $(b-1)$ columns of $P''$ must have Hamming weight $\ge 3.$ Next, we will show that of the remaining columns of $P''$, in any set of $b$ consecutive columns, at most one column can have Hamming weight $2$. Using these facts we will show the sparsity claim for Construction I. We will then prove the field size claim.

We now determine the least number of non-zero entries that $H''$ must have. From similar arguments as that in the proof of Theorem \ref{thm:binary_to_ternary_optimality}, we have $H''(0,j)\ne0$ and $H''(b,j)\ne0,~j\in[b+1:n-1].$ Thus, for $j\in[b+1:n-1],~w_H(H''_j)\ge2.$ 
Now suppose that for some $j\in[b+1:\min\{2b-1,n-1\}],~w_H(H''_j)=2.$ Then, $H''_j$ is in the span of $H''_0$ and $H''_b,$ from which it follows that $\cc$ can not recover from the $(b,1)$-burst whose support is given by $\{0\}\bigcup[b:j].$ Thus, for $j\in[b+1:\min\{2b-1,n-1\}],~w_H(H''_j)\ge3.$
Now suppose that $n>(2b+1),$ and that for some $j\in[2b:n-2]$ there exists $j_1\in[j+1:\min\{j+b-1,n-1\}]$ such that $w_H(H''_j)=w_H(H''_{j_1})=2.$ Then, we can linearly combine $H''_j$ and $H''_{j_1}$ to get a vector $\un{v}\in\textnormal{span}\langle\{H_0\}\rangle.$ However, this implies that $\cc$ can not recover from the $(b,1)$-burst whose support is given by $\{0\}\bigcup[j:j_1],$ which is a contradiction. It follows that out of any $b$ consecutive columns of $H''(:,[2b:n-1]),$ at most one column can have Hamming weight $2,$ and the rest must have Hamming weight at least 3.

Let $H$ be the parity-check matrix of the $[n,n-(b+1)]$ $(b, 1)$-code obtained via Construction I. It is easy to verify that the row operation $H(0,:)\rightarrow(H(0,:)-H(b,:))$
yields a matrix $\hat{H}$ of the form $[I_{b+1}\mid \hat{P}]$. Now observe that the set of columns of $\hat{H}$ that have Hamming weight $2$ is $\{\hat{H}_{jb}\mid j\in[2:l_{(n,b)}-1]\}$, and all the other columns (except the first $(b+1)$ columns) have Hamming weight 3. It follows that $\hat{H}$ has maximum sparsity.

We  will now show the field size claim. Theorem \ref{thm:binary_to_ternary_optimality} and the argument above show that $q^*=2$ for $(b+1)<n\le2b,$ and $q^*=3$ for $2b<n\le3b.$ Now consider $n>3b.$ Suppose that $H''\in\bb{F}_{q}^{{(b+1)\times n}}$ is such that the matrix obtained by reversing the rows and columns of $H''$ is an element of $\mathcal{H}(n,b),$ and $H''$ has the same number of non-zero entries as $\hat{H}$. Then, $H''$ must have $(l_{(n,b)}-2)$ columns having Hamming weight $2,$ say the columns indexed by $\{i_1,i_2,\dots,i_{l_{(n,b)}-2}\}.$ The support of each of these columns is $\{0,b\}.$ We can scale each of these columns so that $H''(0,i_j)=1,~j\in[1:l_{(n,b)}-2],$ without affecting the linear dependence relations among these columns. Since any two columns of $H''$ must be linearly independent, $\{H''(b,i_j)\mid j\in[1:l_{(n,b)}-2]\}$ must be a set of $(l_{(n,b)}-2)$ distinct non-zero elements of $\bb{F}_{q}.$ It follows that $q\ge(l_{(n,b)}-1).$
\epf

\section{An Upper-Bound on the Minimum Distance of Cyclic Codes}
\label{sec:cyc_code_results}
We will now derive an upper-bound on the minimum distance of a cyclic code. To the best of our knowledge, this bound is not known in the literature.
\bthm
\label{thm:cyclic_dmin_bound}
Let $\cc$ by an $[n,k,d]$ cyclic code with reciprocal polynomial $h(X)$. Then,
\beq
\label{eq:dm_bound}
d\le (n-k+1)-Z(h(X)).
\eeq
\ethm
\bpf
Let $h(X)=h_0+h_1X+\dots+h_{k-1}X^{k-1}+h_kX^k$ be the reciprocal polynomial of $\cc$, and let $H$ be the parity-check matrix of $\cc$ of the form \eqref{eq:pcm}. The proof mainly relies on the observation that each row of $H$ (except the top row) is a (right) cyclic shift of the row above it, and therefore  $H([n-k-s:n-k-1],[n-k-s:n-1])$ contains at least one all-zero column, where $s=\min\{Z(h(X)),n-k\}$. Now, if $Z(h(X))\ge (n-k)$ then $H$ contains an all-zero column, which implies $d=1.$ It can be argued from the cyclicity of $\cc$ that this can not be the case. Thus, we must have $Z(h(X))<(n-k).$

If $Z(h(X))=0,$ then \eqref{eq:dm_bound} reduces to the Singleton bound. Now consider $Z(h(X))>0$.
From the above observation, $H([n-k-Z(h(x)):n-k-1],[n-k-Z(h(x)):n-1])$ contains at least one all-zero column. 
Let $j\in[n-k-Z(h(x)):n-1]$ be such that $H_j([n-k-Z(h(x)):n-k-1])=\un{0}_{Z(h(x))}$. Since $h_0\ne 0$ and $H([n-k-Z(h(x)):n-k-1],[0:n-k-Z(h(x))-1])=\un{\un{0}}^{Z(h(x))\times(n-k-Z(h(x)))},$ we have that $H_j\in\textnormal{span}\langle\{H_i\ |\ i\in [0:n-k-Z(h(x))-1]\}\rangle.$ 
Thus, $H$ has a set of $(n-k-Z(h(x))+1)$ linearly dependent columns, which implies \eqref{eq:dm_bound}.
\epf

It is well-known that an $[n,k]$ cyclic code can recover from any burst erasure of length $(n-k)$, irrespective of its minimum distance \cite[Chapter 7]{ecc_Mac_Slo}. However, as we show in the theorem below, a characterization of cyclic codes achieving the bound \eqref{eq:dm_bound} can be found in terms of their ability to simultaneously recover from burst and random erasures.

\bthm
An $[n,k,d]$ cyclic code $\cc$ meets the bound \eqref{eq:dm_bound} iff there exists an erasure pattern comprising of a burst erasure of length $(d-1)$ along with 1 random erasure that the code can not recover from.  
\ethm
\bpf
We first show necessity. Let $\cc$ be an $[n,k,d]$ cyclic code with reciprocal polynomial $h(X)$ and parity-check matrix $H$ of the form \eqref{eq:pcm}. Suppose $\cc$ meets the bound \eqref{eq:dm_bound}. Then, the set of linearly dependent columns identified in the proof of Theorem \ref{thm:cyclic_dmin_bound} corresponds a burst erasure of length $(d-1)$ along with 1 random erasure.

We now argue sufficiency. Sufficiency follows for the case $Z(h(X))=0$ from the properties of MDS codes. Now consider the case $Z(h(X))>0$. Suppose that the $[n,k,d]$ cyclic code $\cc$ can not recover from an erasure pattern comprising of a burst erasure of length $(d-1)$ along with 1 random erasure. Since the code is cyclic, we can assume without loss of generality that the burst erasure corresponds to columns $\{H_i,~i\in[0:d-2]\}$. Suppose that the index of the column corresponding to the random erasure is $j\in[d-1:n-1]$. Since $d\le(n-k+1)-Z(h(X))$
from \eqref{eq:dm_bound}, it follows that $H([d-1:n-k-1],[0:d-2])=\un{\un{0}}^{(n-k-d+1)\times(d-1)}.$ Thus, in order for $\{H_j\}\bigcup\{H_i\ |\ i\in [0:d-2]\}$ to be a linearly dependent set, $H_j([d-1:n-k-1])=\un{0}_{(n-k-d+1)}$ must hold. Hence, $H([d-1:n-k-1],[d-1:n-1])$ contains at least one all-zero column. Following similar arguments as in the proof of Theorem \ref{thm:cyclic_dmin_bound}, it can be shown that the smallest $i\in[0:n-k-1]$ for which $H([i:n-k-1],[i:n-1])$ contains an all-zero column is $i=(n-k-Z(h(X))).$
\epf

\bibliographystyle{IEEEtran}
\bibliography{Streaming}

\appendix

\subsection{Proof of Theorem \ref{thm:sub-linear_field_size}}
Let $H$ be the parity-check matrix of $\cc$ constructed via Construction I. We will now form a partition of the columns of $H$. For $s\in[0:b_1-1],$
let $H_i\in \mathcal{A}_s$ if $i=s\pmod{b_1},~i\in[0:n-1].$ 
Observe that if $H_i\in \mathcal{A}_s,~\text{supp}(H_i)=\{s\}$ if $i\in[0:b_1-1],$ and $\text{supp}(H_i)=\{s,~ b_1+(i \pmod{b_2})\}=\{s,~b_1+(s \pmod{b_2})\}$, if $i\in[b_1:n-1].$
Thus, for any $s_1,s_2\in[0:b_1-1]$ such that $s_1\neq s_2,$ any column in $\mathcal{A}_{s_1}$ is linearly independent of any column in $\mathcal{A}_{s_2
}.$ By construction, any two columns $H_i$ and $H_j$ belonging to the same $\mathcal{A}_s$ are linearly independent and the support of any non-trivial linear combination of $H_i$ and $H_j$ is a non-empty subset of $\{s,~b_1+(s\pmod{b_2})\}$. 

Now consider a burst erasure that erases $b_1$ code symbols, and let $I$ be the set of indices of the erased code symbols. Consider another burst erasure that erases $b_2$ code symbols, and let $J=[j_0:j_0+b_2-1]$ be the set of indices of the code symbols erased by this burst. 
Define $s_v=j_0+v\pmod{b_1},~v\in[0:b_2-1].$ Then, for $v\in[0:b_2-1],$ $H_{j_0+v}\in\mathcal{A}_{s_v}$. Since the second burst erases $b_2\le b_1$ consecutive code symbols, we have that for any two distinct $j_1,j_2\in J,$ if $H_{j_1}\in \mathcal{A}_{s'_1}$ and $H_{j_2}\in A_{s'_2},$ then $s'_1\neq s'_2.$ A similar result holds for any two distinct $i_1,i_2\in I.$ It follows that for every $j\in J,$ there exists exactly one $i\in I$ such that if $H_j\in \mathcal{A}_s,$ then $H_i\in \mathcal{A}_s.$
Let $I'=\{i'_0,i'_1,\dots,i'_{b_2-1}\}\subseteq I$ be such that $H_{j_0+v},H_{i'_v}\in A_{s_v},~v\in[0:b_2-1].$ It is easy to see that if a non-trivial linear combination of $\{H_u\mid u\in I\bigcup J\}$ yields the all-zero vector, then any column in $\{H_u\mid u\in I\setminus I'\}$ can not have a non-zero contribution in this linear combination. Further, for any $v\in[0:b_2-1],$ the support of any non-zero linear combination of $H_{j_0+v}$ and $H_{i'_v}$ is a non-empty subset of $\{s_v,~b_1+(s_v\pmod{b_2})\},$ which is disjoint from any subset of $\bigcup_{v'\in[0:b_2-1]\setminus{v}}\{s_{v'},~b_1+(s_{v'}\pmod{b_2})\}$ (which once again follows from the fact that the second burst erases $b_2$ consecutive code symbols). Thus, $\{H_u\mid u\in I\bigcup J\}$ is a linearly independent set.


Clearly, any $3$ columns of $H$ belonging to the same $A_s$ are linearly dependent. If $n>2b_1,$ then $| \mathcal{A}_1|\ge 3.$ Further, if $n\le2b_1,$ it is easy to verify that the minimum distance of $\cc$ is $4.$

\end{document}